\RequirePackage{snapshot}
\documentclass{PoS}

\usepackage[numbers,square,sort&compress]{natbib}

\title{Vertexing and Tracking Software at Belle II}

\ShortTitle{Vertexing and Tracking Software at Belle II}

\author{Tobias Schl{\"u}ter%
  \thanks{This research was supported by the DFG cluster of excellence
    'Origin and Structure of the Universe' and
    under BMBF Contract 05H12WM8.}\hspace{2mm} for the Belle II Software Group\\
  Ludwig-Maximilians-Universit{\"a}t\\
  E-mail: \email{tobias.schlueter@physik.uni-muenchen.de}}

\abstract{ Belle~II is a $B$ factory experiment aiming to start
  physics data taking in 2017.  It is currently being set up at the
  SuperKEKB accelerator at the KEK facility in Tsukuba (Japan), an
  asymmetric $e^+e^-$ collider which aims to achieve an unprecedented
  instantaneous luminosity of $8\cdot10^{35}
  \textrm{Hz}/\textrm{cm}^2$.  This forty-fold increase over
  predecessor experiments is achieved by employing a novel nano-beam
  scheme.  Originally developed for the now-defunct SuperB experiment,
  this scheme allows a significant increase in luminosity at only
  modest increases of beam currents.  Challenges for the vertex
  detector result from increased data and background rates.  At full
  luminosity, physics data will be recorded at a rate of
  $30\,\textrm{kHz}$.  The radiation-hard DEPFET-sensors of the
  innermost layer of the vertex detector will be read out employing a
  novel data-reduction scheme using selective detector read out based
  on online reconstruction of event data.  Belle~II uses a software
  framework in which data handling is unified between various data
  processing modules.  In this way, tasks can be divided flexibly and
  the same software framework can be used for a diversity of tasks
  ranging from the high-level trigger to final processing of plots for
  publications.  Tracking and vertexing modules are currently under
  development and we will discuss their features.}

\FullConference{The 23rd International Workshop on Vertex Detectors\\
		 15-19 September 2014\\
		 Macha Lake, The Czech Republic}


\begin{document}

\newcommand{\belle}{Belle~II}
\newcommand{\genfit}{\textsc{Genfit}}

\section{Introduction}

Our experimental knowledge of the flavor structure of the Standard
Model is in large parts due to the $B$-factories Belle and BaBar.
Situated at the asymmetric $e^+e^-$ colliders KEKB and PEP-II,
respectively, they collected data throughout the first decade of the
2000s, confirming in great detail the CKM theory, and the related
violation of the fundamental discrete symmetries $C$, $P$ and
$T$~\cite{Bevan:2014iga}.  \belle{} is a successor to these
experiments~\cite{Abe:2010gxa}.  It employs the same concept of
asymmetric $e^+e^-$ collisions at the $\Upsilon(10580)$ resonance
which decays into two mesons $B\bar B$, but compared to the
predecessor experiments \belle{} uses a slightly reduced beam momentum
asymmetry.  Typical travel distances of the $B$ mesons before their
decays are $200\,\mu\textrm{m}$.  \belle{} is an exploratory
experiment aiming to test the predictions of the flavor sector of the
Standard Model at unprecedented precision, thus probing new physics
contributions well into the $\textrm{TeV}$ mass range, given
sufficient misalignment of the new physics flavor structure with that
of the Standard Model~\cite{Aushev:2010bq}.  The \belle{} experiment
is currently assembled at the SuperKEKB accelerator at the KEK
facility in Tsukuba, Japan.  The experiment is organized as an
international collaboration of some 600 researchers and will start
taking physics data in 2017, quickly integrating more luminosity than
its predecessor experiments, then aiming for a total integrated
luminosity $50\,\textrm{ab}^{-1}$ accumulated over a period of
approximately eight years.

The target luminosity of $8\cdot10^{35} \textrm{Hz}/\textrm{cm}^2$ is
approximately forty times larger than that of the predecessor facility
KEKB.  This is achieved by a novel nano-beam scheme for the
interaction region supplemented by a two-fold increase in beam
currents~\cite{Ohnishi:2013fma}.  The increased luminosity leads to a
corresponding increase in background levels, mainly QED backgrounds
(which scale proportionally to luminosity), and in-beam scattering
(Touschek effect, due to stronger focussing).  Other machine
backgrounds only scale as the beam current.

\section{The \belle{} detector}

The \belle{} detector is an upgrade of the Belle detector: it is
housed in the same structure, reusing the spectrometer magnet.  The
design aimed for performance comparable to the performance of the
Belle detector in spite of the increased background levels expected at
the higher-luminosity accelerator.  Where most parts of the detector
were refined in various ways, two major new components were introduced
in the design: a pixel vertex detector based on the DEPFET technology
very close to the interaction point which is capable of withstanding
the increased radiation levels while providing excellent resolution,
and a complete overhaul of the particle identification detectors which
are of Cherenkov imaging type.  Additionally, the beam pipe is
replaced with a smaller, thinner device, which allows installing the
innermost layer of the vertex detector at a distance of only
$1.4\,\textrm{cm}$ from the interaction point.

The components of the \belle{} vertex detector are discussed in detail
in Peter Kodys's~\cite{Kodys:2014aa}, Michael
Schnell's~\cite{Schnell:2014aa} and Lorenzo
Vitale's~\cite{Vitale:2014aa} contributions to this conference, so we
will restrict ourselves to an executive summary.

\belle{} is designed for detecting and reconstructing particle
trajectories for transverse momenta exceeding $50\,\textrm{MeV}/c$
while providing excellent momentum resolution all the way up to $p_T$
at the physical limit which is half the center of mass energy
$m(\Upsilon(4S))=10.58\,\textrm{GeV}/c^2$.  The vertex detector
consists of six layers of silicon detectors, situated at radial
distances from the interaction point between $1.4\,\textrm{cm}$ and
$14\,\textrm{cm}$.  It is surrounded by a wire-chamber that extends up
to a radius of $1.1\,\textrm{m}$.

\begin{figure}[!t]
  \centering
  \includegraphics[width=\linewidth]{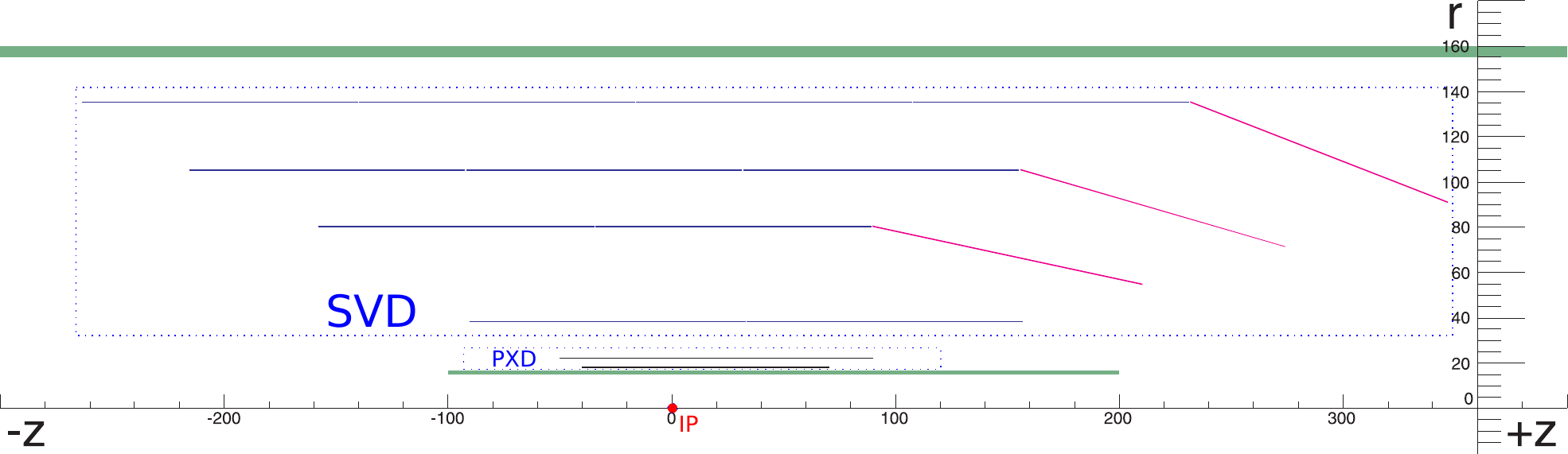}
  \caption{Layout of the \belle{} vertex detector (VXD).  Distances
    are indicated in mm relative to the interaction point (IP).  The
    two layers of the pixel detector (PXD) and the four layers of the
    silicon vertex detector (SVD) are indicated.  The forward parts of
    the outer three layers of the SVD are slanted to reduce the amount
    of silicon.  The VXD volume is bounded by the beam pipe on the
    inside and by the inner wall of the drift chamber on the outside
    (both indicated in green).}
  \label{fig:vxd-layout}
\end{figure}

\begin{figure}[htbp]
  \centering
  \includegraphics[width=\textwidth]{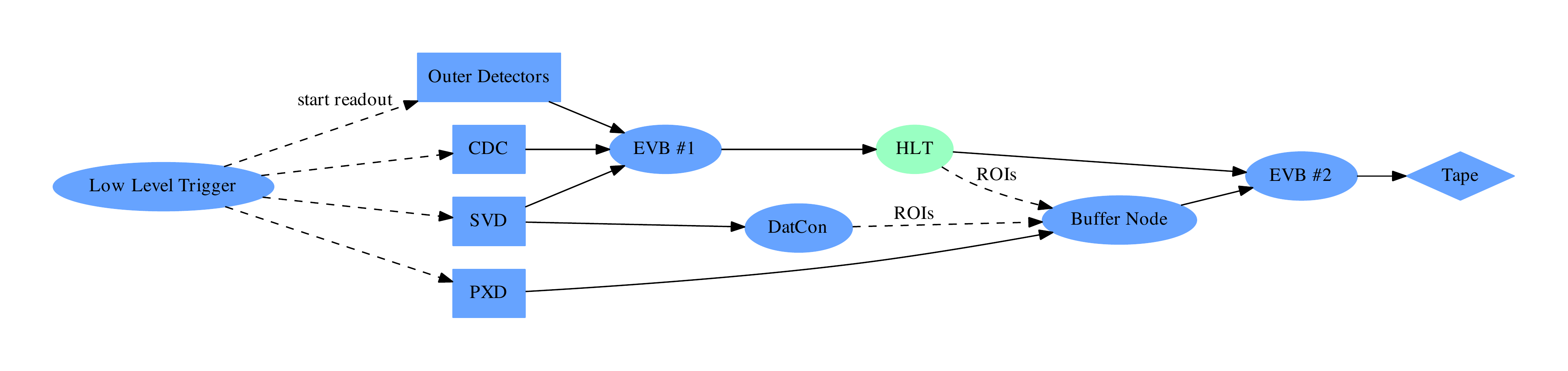}
  \caption{The two-step event-building procedure of the \belle{}
    experiment.  The Low Level Trigger causes readout of the various
    subdetectors.  Data from the pixel detector (PXD) is stored on a
    buffer node, the data from the wire-chamber (CDC), the silicon
    strip detector (SVD) and the other detectors is combined on the
    first event builder (EVB \#1).  The High Level Trigger (HLT)
    processes the data, and transmits regions-of-interest to the
    buffer node, which then passes the relevant data from the pixel
    detector to the second event Builder (EVB \#2), which combines the
    pixel data in the regions-of-interest with the data from the other
    detectors thus creating the final data samples that are stored on
    disk.  An alternative approach to the definition of
    regions-of-interest is implemented in the DatCon hardware which
    employs Hough-transform-based reconstruction of the SVD data
    implemented on FPGA hardware.}
  \label{fig:readout}
\end{figure}

The vertex detector itself consists of two subdetectors, cf.\
Fig.~\ref{fig:vxd-layout}.  The two innermost layers are built from
DEPFET-type active sensors with pixel readout~\cite{Alonso:2012ss},
able to withstand the high radiation levels.  These semiconductor
detectors are thinned to $75\,\mu\textrm{m}$ in the active area
corresponding to a radiation length of $X/X_0=0.21\%$.  The
surrounding four layers are built using double-sided silicon strip
detectors of thickness $300\,\mu\textrm{m}$, read out using APV25
chips~\cite{Friedl:2013gta}.  Readout has to deal with the long
integration times of the DEPFET sensors ($\approx 20\,\mu\textrm{s}$)
which is comparable to the physics trigger rate ($30\,\textrm{kHz}$).
Given the expected occupancy in the range of $1\%$, full readout at
every trigger would lead to a data rate of $20\,\textrm{GB}/s$ for the
pixel layers alone, exceeding the amount of data collected by the
remainder of \belle{} by an order of magnitude.  In order to reduce
this to something supportable, an online data reduction scheme has to
be put in place.  This is implemented by a two-level event
building/high-level triggering scheme, illustrated in
Fig.~\ref{fig:readout}.  Upon occurence of a low-level trigger signal,
all subdetectors besides the pixel detector are read out, events are
built and fed to the high-level trigger.  Simultaneously, the
pixel-detector data is read out and stored on a buffer
node~\cite{Gessler:2014gba}.  The high-level trigger performs a full
reconstruction of the event sans pixel data.  Reconstructed tracks are
then extrapolated back to the pixel layers, where the vicinities of
the estimated crossing points define regions-of-interest which are read
from the buffer node and added to the event in a second event-building
step~\cite{Bilka:2014lla}.  As an alternative, an FPGA-based
implementation of this region-of-interest definition is implemented on
dedicated readout hardware of the silicon strip detector.  This
approach was discussed in another contribution to this
conference~\cite{Schnell:2014aa}.  In order to recover tracks with
very low transverse momenta, additionally pixel data corresponding to
very large ionisation is recorded irrespective of whether it can be
associated with a region-of-interest.  In this way a ten-fold
reduction of the amount of data recorded can be achieved.

The wire chamber construction was recently completed.  The detector is
now being tested and first calibrations using cosmic rays are being
performed.  It has a total of 14k sense wires and 42k field wires,
distributed over the $8000\,\ell$ volume.  The signal wires are
organized in 56 radial layers which are subdivided into nine
super-layers, of which five have wires along the direction of the
magnetic field (``axial wires'') and four are inclined in order to
determine the track coordinates in the direction of the field
(``stereo wires'')~\cite{Taniguchi:2013pwa}.

\section{The \belle{} software framework}

\belle{} unifies all data processing tasks in a single software
package, {\tt
  basf2}~\cite{Itoh:2012hb,Lee:2011za,Moll:2011zz,Kim:2014aa} which
runs on standard Linux systems.  This package consists of a flexible
framework of individual modules which process the data.  The modules
are organized in a user-defined chain.  Each module can read data
provided by the preceding modules via the so-called data store, and it
can also add additional data to the data store.  Special modules can
read data from disk or from detectors, or send and receive data over
the network.  As an illustration of this, consider a track-finding
module: it will read detector hits from the data store and write track
candidates to the data store.  If it processes MC data, it may also
read a relation between the hits and MC truth information, and use
this to establish a relation between track candidates and the
particles created during the simulation.  These track candidates can
then be processed by a track fitting module.

Data store contents can be serialized or read at any point in the
chain.  This functionality is used in several ways.  Besides storage
of the processed data, this allows distributing the data over multiple
cores or networked hosts transparently, and it allows an arbitrary
splitting of the taks between different computer jobs.  E.g.\ an
alternative implementation of a module can be tested without repeating
any of the work done before: one processes the chain up to, but not
including, the module under test, and stores the result to disk.  Now
any change to the module can be tested immediately, by reading the
data from disk and processing the chain starting from the module under
inspection.

\begin{figure}[htbp]
  \centering
  \includegraphics[width=\linewidth]{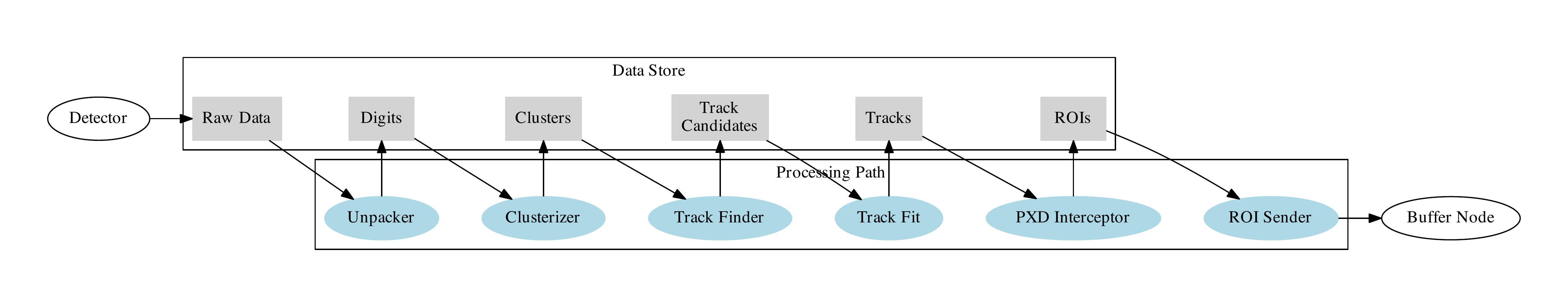}
  \caption{Example block-diagram visualization of the data flow in the
    High-Level Trigger processing chain.  Each module is indicated as
    a blue oval.  Grey rectangles indicate the data store arrays.  The
    flow of data is indicated by the arrows.}
  \label{fig:dataflow}
\end{figure}

The design is driven in particular by the needs of the high-level
trigger of the \belle{} experiment, which uses a networked computing
farm running the full reconstruction software, and which plays an
essential role in the readout of the pixel vertex detector, as
explained above.  As a complement to Fig.~\ref{fig:readout}, we show
as an example of such a processing path the sequence of modules used
in a recent beam test where pixel and silicon strip layers as well as
the full readout architecture including the high-level trigger were
employed~\cite{Bilka:2014lla}.  The sequence shown in
Fig.~\ref{fig:dataflow} shows how the incoming raw detector data is
sequentially processed to yield the regions-of-interest, sent to the
buffer node by the ``ROI Sender'' module in the end.  In the test-beam
experiment, this chain ran on the high-level trigger node, indicated
by HLT in Fig.~\ref{fig:readout}.

\section{Tracking and vertexing software}

Physics analysis in \belle{} will concern itself with the ongoings
near the interaction point, the objects of interest in a typical
analysis are the decay products of the $B$ and $\bar B$ mesons
produced in the initial reactions.  The $B$ mesons decay in close
vicinity of a few hundred $\mu\textrm{m}$ to the interaction point
which in the employed nanobeam scheme is itself very well-defined.  In
this vein, the reconstructed objects used for analysis are
parameterized in the vicinity of the interaction point.  In the mDST
data format used for analysis, track parameters are stored in a
perigee helix parametrization together with information from particle
identification.  $V^0$ objects which decay at appreciable distances
from the interaction point are reconstructed from their decay tracks
before being stored in the mDST files.  Photons and $K^0_L$ are stored
using the information from the respective detectors.  This way, the
original $B$ decay vertices can be reconstructed efficiently on the
level of physics analysis and file sizes are minimized.  In the
following we will discuss the reconstruction steps that lead to the
mDST data.

\subsection{Track finding}

Track candidates are identified independently in the vertex detector
and the wire chamber.  In the vertex detector, a cellular-automaton
algorithm is used to cope with the potentially large combinatorics,
especially in the presence of
background~\cite{FruHwirth:2013jta,Bilka:2014lla}.  Adjacent pairs of
hits are combined into what is called a cell.  If two different cells
share a hit, they become neighbors in the space of the cellular
automaton.  The cellular automaton is then evolved where the cell
state develops according to the number of geometrically allowed
neighbors and the state of the neighbors.  From this procedure results
a set of possible combinations of hits, where the same hits may appear
in several combinations.  A Hopfield network, trained with Monte Carlo
simulations, is then used to extract the best non-overlapping set of
track candidates, which are forwarded to the track fit.

In the wire chamber, besides an adapted version of the track finding
algorithm used in the predecessor experiment Belle, two newly
developed track finders are used: a global track finder based on
Legendre transformations which is limited to tracks emitted in the
vicinity of the interaction point, and a local track finder which uses
a cellular automaton to follow tracks through the wire chamber volume.

After track fitting, compatible tracks from both subdetectors are
merged, and tracks for which no compatible track in the other
subdetector could be found are extrapolated, looking for unassociated
hits which can be added.  The stored tracks are updated to reflect
these operations.

\subsection{Track fitting}

The track candidates are then fitted using a largely overhauled
version of the {\sc Genfit} track fitting
package~\cite{Hoppner:2009af,Rauch:2014wta}.  On the high-level
trigger, the default tracking algorithm is the usual Kalman
fitter~\cite{Fruhwirth:1987fm}.  On the other hand, in physics
analysis the adaptive deterministic annealing
algorithm~\cite{Fruhwirth:2006ph} is used.  It is an extension of the
Kalman fitter that uses an annealing procedure to reject
background-induced hits that were mistakenly added by the track
finding algorithms, and it can also resolve left/right ambiguities in
the wire chamber measurements.  For the purpose of alignment, tracks
are reconstructed with the General Broken Lines
algorithm~\cite{Kleinwort:2012np}, which is a global formulation of
the Kalman fit with a discretized model of material.  The alignment
procedure is described in another contribution to this
conference~\cite{Bilka:2014aa}.

The fitted tracks are extrapolated outwards to the various particle
identification detectors (Cherenkov detectors, electromagnetic
calorimeter, and muon/$K^0_L$ id chambers).  The ionization deposits
in the silicon detector layers and the wire chamber are also used for
$\mathrm{d}E/\mathrm{d}x$-based particle identification.  The
Cherenkov detector places stringent limits on the precision of the
time-of-flight determination which is evaluated during the track fit
based on the individual segments between the individual hits.  The
tracks are extrapolated towards the interaction point, and the helix
information in the perigee parametrization is stored in the mDST.

\subsection{Vertex finding and fitting}

In order to reconstruct $V_0$ type decays, all negative tracks are
paired up with all positive tracks, and a vertex fit is performed
using the standard Kalman fit implemented in the RAVE
package~\cite{Waltenberger:2011zz}.  If a vertex that is separated
from the interaction region could be reconstructed by this procedure,
a $V_0$ object is added to the mDST.  Physics analysis then uses the
reconstructed tracks, the $V_0$s and additional calorimeter and
$K^0_L$ objects stored in the mDSTs to recover the $B$ mesons.  The
$B$ vertex is reconstructed using either the RAVE package or the
vertexing tool adapted from the Belle experiment, KFIT.  As a
promising new development, the increased data sample of the \belle{}
experiment allows to perform with unprecedented statistics analyses
where all decay particles of at least one of the $B$, $\bar B$ mesons
are reconstructed, and thus the complete event (up to invisible
particles from one of the decays) can be reconstructed.  The small
size of the beam crossing region allows additional strong constraints
as the transverse momentum of the $B$ mesons is strongly correlated
with the observed displacement from the interaction point.

\section{Outlook}

The \belle{} experiment will begin its physics data taking in 2017.
Owing to its significantly increased luminosity, it will allow great
improvements in measurements that were thus far limited by statistics.
New developments such as the full-event reconstruction will allow to
put stringent limits on invisible $B$ decays or decays with only one
charged particle.  Not discussed in this proceeding, but worth
mentioning are also the large samples of $D\bar D$ and $\tau^+\tau^-$
events whose analysis will also benefit from \belle{}'s advanced
vertexing capabilities.


\end{document}